# Voice Recognition Algorithms using Mel Frequency Cepstral Coefficient (MFCC) and Dynamic Time Warping (DTW) Techniques

Lindasalwa Muda, Mumtaj Begam and I. Elamvazuthi

**Abstract**— Digital processing of speech signal and voice recognition algorithm is very important for fast and accurate automatic voice recognition technology. The voice is a signal of infinite information. A direct analysis and synthesizing the complex voice signal is due to too much information contained in the signal. Therefore the digital signal processes such as Feature Extraction and Feature Matching are introduced to represent the voice signal. Several methods such as Liner Predictive Predictive Coding (LPC), Hidden Markov Model (HMM), Artificial Neural Network (ANN) and etc are evaluated with a view to identify a straight forward and effective method for voice signal. The extraction and matching process is implemented right after the Pre Processing or filtering signal is performed. The non-parametric method for modelling the human auditory perception system, Mel Frequency Cepstral Coefficients (MFCCs) are utilize as extraction techniques. The non linear sequence alignment known as Dynamic Time Warping (DTW) introduced by Sakoe Chiba has been used as features matching techniques. Since it's obvious that the voice signal tends to have different temporal rate, the alignment is important to produce the better performance.This paper present the viability of MFCC to extract features and DTW to compare the test patterns.

**Index Terms**— Feature Extraction, Feature Matching, Mel Frequency Cepstral Coefficient (MFCC), dynamic Time Warping (DTW)

——————————— ◆ ———————————

## 1 INTRODUCTION

VOICE Signal Identification consist of the process to convert a speech waveform into features that are useful for further processing. There are many algorithms and techniques are use. It depends on features capability to capture time frequency and energy into set of coefficients for cepstrum analysis. [1].

Generally, human voice conveys much information such as gender, emotion and identity of the speaker. The objective of voice recognition is to determine which speaker is present based on the individual's utterance [2].Several techniques have been proposed for reducing the mismatch between the testing and training environments. Many of these methods operate either in spectral [3,4], or in cepstral domain [5]. Firstly, human voice is converted into digital signal form to produce digital data representing each level of signal at every discrete time step. The digitized speech samples are then processed using MFCC to produce voice features. After that, the coefficient of voice features can go trough DTW to select the pattern that matches the database and input frame in order to minimize the resulting error between them.

The popularly used cepstrum based methods to compare the pattern to find their similarity are the MFCC and DTW. The MFCC and DTW features techniques can be implemented using MATLAB [6]. This paper reports the findings of the voice recognition study using the MFCC and DTW techniques.

The rest of the paper is organized as follows: principles of voice recognition is given in section 2, the methodology of the study is provided in section 3, which is followed by result and discussion in section 4, and finally concluding remarks are given in section 5.

## 2 PRINCIPLE OF VOICE RECOGNITION

### 2.1 Voice Recognition Algorithms

A voice analysis is done after taking an input through microphone from a user. The design of the system involves manipulation of the input audio signal. At different levels, different operations are performed on the input signal such as Pre-emphasis, Framing, Windowing, Mel Cepstrum analysis and Recognition (Matching) of the spoken word.

The voice algorithms consist of two distinguished phases. The first one is training sessions, whilst, the second one is referred to as operation session or testing phase as described in figure 1 [7].

————————————————

- *Lindasalwa Muda is with the Department of Electrical and Electronic Engineering, Universiti Teknologi PETRONAS Bandar Seri Iskandar 31750 Tronoh.Perak, MALAYSIA.*

- *Mumtaj Begam is with the Department of Electrical and Electronic Engineering, Universiti Teknologi PETRONAS Bandar Seri Iskandar 31750 Tronoh.Perak, MALAYSIA.*

- *I. Elamvazuthi is with the Department of Electrical and Electronic Engineering, Universiti Teknologi PETRONAS Bandar Seri Iskandar 31750 Tronoh.Perak, MALAYSIA.*

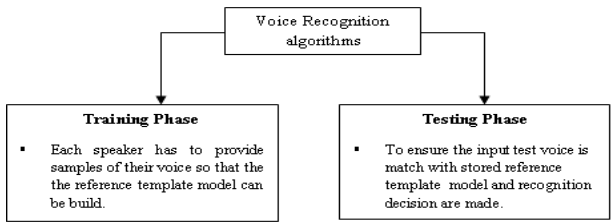

Fig. 1. Voice Recognition algorithms

## 2.2 Feature Extraction (MFCC)

The extraction of the best parametric representation of acoustic signals is an important task to produce a better recognition performance. The efficiency of this phase is important for the next phase since it affects its behavior.

MFCC is based on human hearing perceptions which cannot perceive frequencies over 1Khz. In other words, in MFCC is based on known variation of the human ear's critical bandwidth with frequency [8-10]. MFCC has two types of filter which are spaced linearly at low frequency below 1000 Hz and logarithmic spacing above 1000Hz. A subjective pitch is present on Mel Frequency Scale to capture important characteristic of phonetic in speech.

The overall process of the MFCC is shown in Figure 2 [6, 7]:

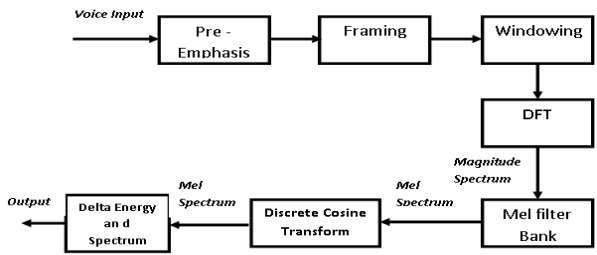

Fig. 2. MFCC Block Diagram [6,7]

As shown in Figure 3, MFCC consists of seven computational steps. Each step has its function and mathematical approaches as discussed briefly in the following:

**Step 1:** Pre–emphasis
This step processes the passing of signal through a filter which emphasizes higher frequencies. This process will increase the energy of signal at higher frequency.

$$Y[n] = X[n] - 0.95 X[n-1] \quad (1)$$

Lets consider a = 0.95, which make 95% of any one sample is presumed to originate from previous sample.

**Step 2:** Framing
The process of segmenting the speech samples obtained from analog to digital conversion (ADC) into a small frame with the length within in the range of 20 to 40 msec. The voice signal is divided into frames of N samples. Adjacent frames are being separated by M (M<N). Typical values used are M = 100 and N= 256.

**Step 3:** Hamming windowing

Hamming window is used as window shape by considering the next block in feature extraction processing chain and integrates all the closest frequency lines. The Hamming window equation is given as:
If the window is defined as W (n), 0 ≤ n ≤ N-1 where

N = number of samples in each frame
Y[n] = Output signal
X (n) = input signal
W (n) = Hamming window, then the result of windowing signal is shown below:

$$Y(n) = X(n) \times W(n) \quad (2)$$

$$w(n) = 0.54 - 0.46 \cos\left(\frac{2\pi n}{N-1}\right) \quad 0 \le n \le N-1 \quad (3)$$

**Step 4:** Fast Fourier Transform
To convert each frame of N samples from time domain into frequency domain. The Fourier Transform is to convert the convolution of the glottal pulse U[n] and the vocal tract impulse response H[n] in the time domain. This statement supports the equation below:

$$Y(w) = FFT[h(t) * X(t)] = H(w) * X(w) \quad (4)$$

If X (w), H (w) and Y (w) are the Fourier Transform of X (t), H (t) and Y (t) respectively.

**Step 5:** Mel Filter Bank Processing
The frequencies range in FFT spectrum is very wide and voice signal does not follow the linear scale. The bank of filters according to Mel scale as shown in figure 4 is then performed.

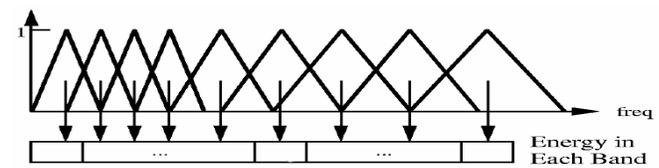

Fig. 3. Mel scale filter bank, from (young et al,1997)

This figure shows a set of triangular filters that are used to compute a weighted sum of filter spectral components so that the output of process approximates to a Mel scale. Each filter's magnitude frequency response is triangular in shape and equal to unity at the centre frequency and decrease linearly to zero at centre frequency of two adjacent filters [7, 8]. Then, each filter output is the sum of its filtered spectral components. After that the following equation is used to compute the Mel for given frequency f in HZ:

$$F(Mel) = [2595 * \log 10[1 + f]700] \quad (5)$$

**Step 6:** Discrete Cosine Transform
This is the process to convert the log Mel spectrum into time domain using Discrete Cosine Transform (DCT). The result of the conversion is called Mel Frequency Cepstrum Coefficient. The set of coefficient is called acoustic vectors. Therefore, each input utterance is transformed into a sequence of acoustic vector.

**Step 7:** Delta Energy and Delta Spectrum

The voice signal and the frames changes, such as the slope of a formant at its transitions. Therefore, there is a need to add features related to the change in cepstral features over time . 13 delta or velocity features (12 cepstral features plus energy), and 39 features a double delta or acceleration feature are added. The energy in a frame for a signal x in a window from time sample t1 to time sample t2, is represented at the equation below:

$$Energy = \sum X^2[t] \qquad (6)$$

Each of the 13 delta features represents the change between frames in the equation 8 corresponding cepstral or energy feature, while each of the 39 double delta features represents the change between frames in the corresponding delta features.

$$d(t) = \frac{c(t+1) - c(t-1)}{2} \qquad (7)$$

## 2.3 Feature Matching (DTW)

DTW algorithm is based on Dynamic Programming techniques as describes in [11]. This algorithm is for measuring similarity between two time series which may vary in time or speed. This technique also used to find the optimal alignment between two times series if one time series may be "warped" non-linearly by stretching or shrinking it along its time axis. This warping between two time series can then be used to find corresponding regions between the two time series or to determine the similarity between the two time series. Figure 4 shows the example of how one times series is 'warped' to another [12].

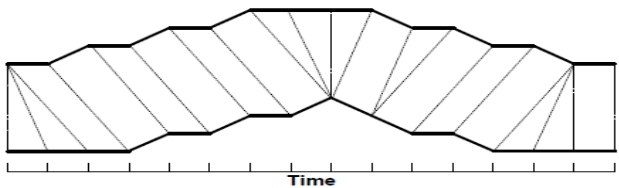

Fig. 4. A Warping between two time series [12]

In Figure 4, each vertical line connects a point in one time series to its correspondingly similar point in the other time series. The lines have similar values on the *y*-axis, but have been separated so the vertical lines between them can be viewed more easily. If both of the time series in figure 4 were identical, all of the lines would be straight vertical lines because no warping would be necessary to 'line up' the two time series. The warp path distance is a measure of the difference between the two time series after they have been warped together, which is measured by the sum of the distances between each pair of points connected by the vertical lines in Figure 4. Thus, two time series that are identical except for localized stretching of the time axis will have DTW distances of zero. The principle of DTW is to compare two dynamic patterns and measure its similarity by calculating a minimum distance between them. The classic DTW is computed as below [13]:

Suppose we have two time series $Q$ and $C$, of length $n$ and $m$ respectively, where:

$$Q = q1, q2,\ldots, qi,\ldots,qn \qquad (1)$$
$$C = c1, c2,\ldots, cj,\ldots,cm \qquad (2)$$

To align two sequences using DTW, an *n*-by-*m* matrix where the (*i*th, jth) element of the matrix contains the distance $d(qi, cj)$ between the two points $qi$ and $cj$ is constructed. Then, the absolute distance between the values of two sequences is calculated using the Euclidean distance computation:

$$d(qi,cj) = (qi - cj)^2 \qquad (3)$$

Each matrix element $(i, j)$ corresponds to the alignment between the points $qi$ and $cj$. Then, accumulated distance is measured by:

$$D(i,j) = \min[D(i-1, j-1), D(i-1, j), D(i, j-1)] + d(i,j) \qquad (4)$$

This is shown in Figure 5 where the horizontal axis represents the time of test input signal, and the vertical axis represents the time sequence of the reference template. The path shown results in the minimum distance between the input and template signal. The shaded in area represents the search space for the input time to template time mapping function. Any monotonically non decreasing path within the space is an alternative to be considered. Using dynamic programming techniques, the search for the minimum distance path can be done in polynomial time P(t), using equation below[14]:

$$P(t) = O\left(N^2 V\right) \qquad (5)$$

where, N is the length of the sequence, and V is the number of templates to be considered.

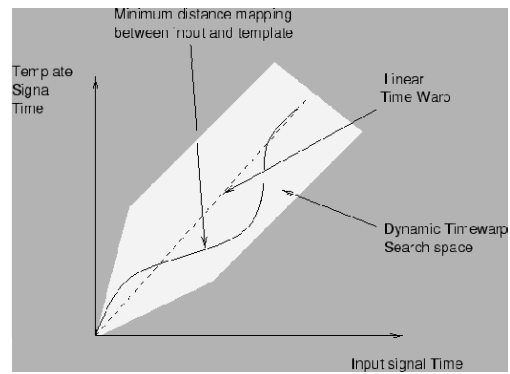

Fig.5. Example Dynamic time warping (DTW) [15]

Theoretically, the major optimizations to the DTW algorithm arise from observations on the nature of good paths through the grid. These are outlined in Sakoe and Chiba [16] and can be summarized as:

*Monotonic condition*: the path will not turn back on itself, both i and j indexes either stay the same or increase, they never decrease.

*Continuity condition*: The path advances one step at a time. Both i and j can only increase by 1 on each step along the path.

*Boundary condition*: the path starts at the bottom left and ends at the top right.

*Adjustment window condition*: a good path is unlikely to wander very far from the diagonal. The distance that the path is allowed to wander is the window length *r*.

*Slope constraint condition*: The path should not be too steep or too shallow. This prevents very short sequences matching very long ones. The condition is expressed as a ratio n/m where m is the number of steps in the x direction and m is the number in the y direction. After m steps in x you must make a step in y and vice versa.

## 3 METHODOLOGY

As mentioned in [12], voice recognition works based on the premise that a person voice exhibits characteristics are unique to different speaker. The signal during training and testing session can be greatly different due to many factors such as people voice change with time, health condition (e.g. the speaker has a cold), speaking rate and also acoustical noise and variation recording environment via microphone. Table II gives detail information of recording and training session, whilst Figure 6 shows the flowchart for overall voice recognition process.

**Table 1. Training Requirement**

| Process | Description |
|---|---|
| 1) Speaker | One Female |
|  | One Male |
| 2) Tools | Mono microphone |
|  | Gold Wave software |
| 3) Environment | Laboratory |
| 4) Utterance | Twice each of the following words: |
|  | On TV |
|  | Off TV |
|  | Volume Up |
|  | Volume Down |
|  | Channel One |
| 5) Sampling Frequency, fs | 16000Khz |
| 6) Feature computational | 39 double delta MFCC coefficient. |

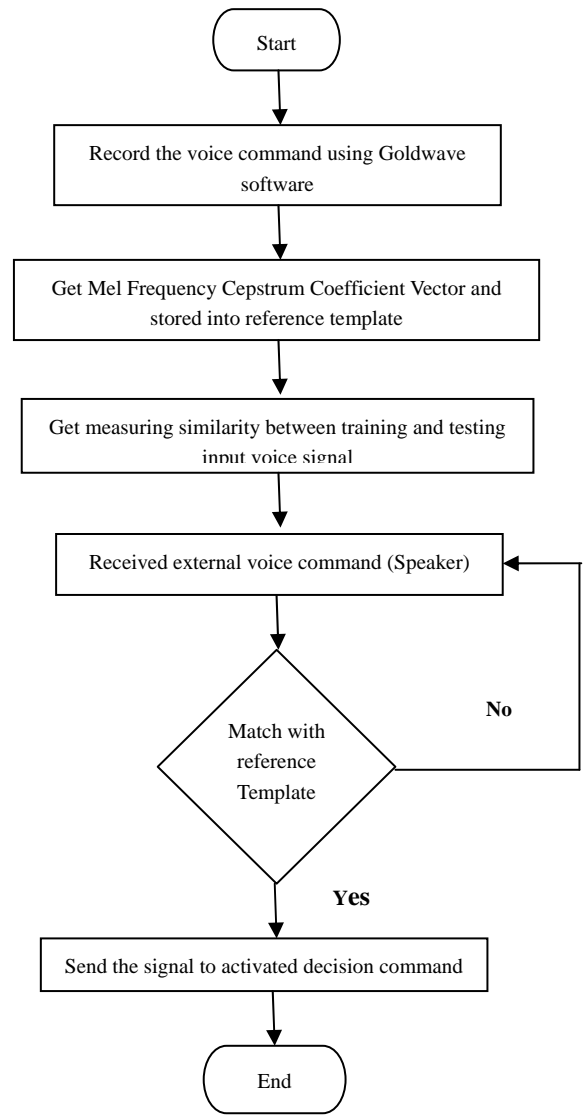

Fig.6. Voice Algorithm flow Chart

## 4 RESULT AND DISCUSSION

The input voice signals of two different speakers are shown in Figure 7.

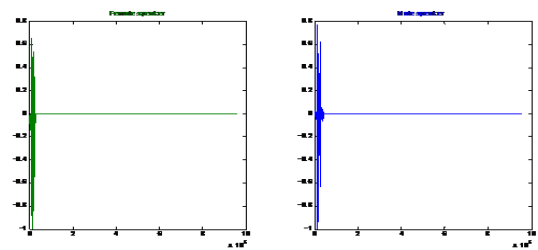

Fig.7. Example voice signal input of two difference speakers

Figure 7 is used for carrying the voice analysis performance evaluation using MFFC. A MFCC cepstral is a matrix, the problem with this approach is that if constant window spacing is used, the lengths of the input and stored sequences is unlikely to be the same. Moreover, within a word, there will be variation in the length of individual phonemes as discussed before, Example the word Volume Up might be uttered with a long /O/ and short final /U/ or with a short /O/ and long /U/.

Figure 8 shows the MFCC output of two different speakers. The matching process needs to compensate for length differences and take account of the non-linear nature of the length differences within the words.

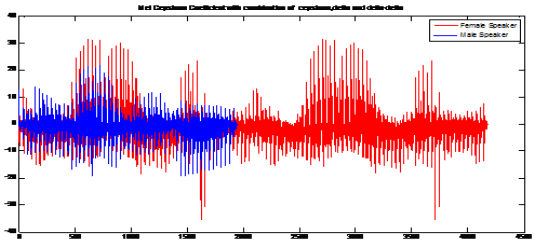

Fig.8. Mel Frequency Cepstrum Coefficients (MFCC) of one Female and Male speaker

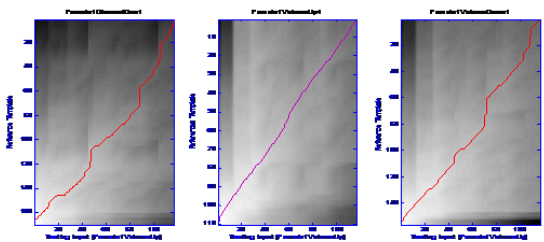

Fig.9. Optimal Warping Path of Test input Female speaker Volume Up

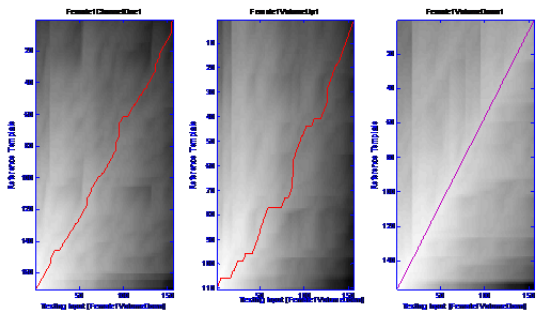

Fig.10. Optimal Warping Path of Test input Female Speaker Volume Down

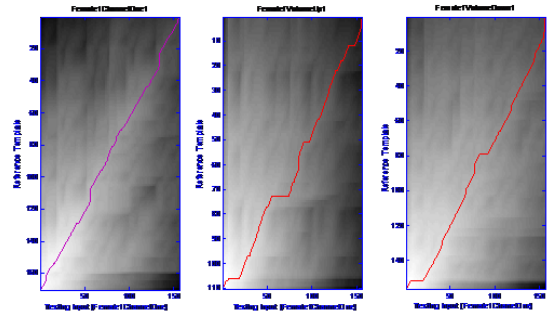

Fig.11. Optimal Warping Path of Test input Female Speaker Channel One

The result shown in Figure 9, 10 and 11 confirms that the input test voice matched optimally with the reference template which was stored in the database. The finding of this study is consistent with the principles of voice recognition outlined in section II where comparison of the template with incoming voice was achieved via a pair wise comparison of the feature vectors in each.

As discussed by [16], the total distance between the sequences is the sum or the mean of the individual distances between feature vectors. The purpose of DTW is to produce warping function that minimizes the total distance between the respective points of the signal. Furthermore, the accumulated distance matrix is used to develop mapping paths which travel through the cells with smallest accumulated distances, then the total distance difference between these two signals is minimized. Through this study, optimal warping path was achieved where the test input matched with the reference template as indicated by the path shown in figures 9 – 11. These findings are consistent with the theory of dynamic time warping as indiacted in Figure 5.

## 5 CONCLUSION

This paper has discussed two voice recognition algorithms which are important in improving the voice recognition performance. The technique was able to authenticate the particular speaker based on the individual information that was included in the voice signal. The results show that these techniques could used effectively for voice recognition purposes. Several other techniques such as Liner Predictive Predictive Coding (LPC), Hidden Markov Model (HMM), Artificial Neural Network (ANN) are currently being investigated. The findings will be presented in future publications.

## ACKNOWLEDGMENT

The authors would like to thank Universiti Teknologi PETRONAS for supporting this work.

**Lindasalwa Binti Muda i**s with the Department of Electrical and Electronic Engineering of Universiti Teknologi PETRONAS (UTP), Malaysia. His research interests include Voice Recognition and Digital signal Processing.

**Mumtaj Begam**  is with the Department of Electrical and Electronic Engineering of Universiti Teknologi PETRONAS (UTP), Malaysia.

**I. Elamvazuthi** is with the Department of Electrical and Electronic Engineering of Universiti Teknologi PETRONAS (UTP), Malaysia.